\documentstyle[preprint,aps]{revtex}

\tightenlines

\begin{document}
\draft
\title{Multi-phonon Raman scattering in semiconductor nanocrystals:
importance of non-adiabatic transitions}
\author{E. P. Pokatilov$^{\ast }$, S. N. Klimin$^{\ast }$, V. M. Fomin$%
^{\ast }$ and J. T. Devreese}
\address{Theoretische Fysica van de Vaste Stof, Universiteit Antwerpen
(U.I.A.), B-2610 Antwerpen, Belgium}
\author{F. W. Wise}
\address{Department of Applied Physics, Cornell University, Ithaca, New York
14853}
\date{November 15, 2001}
\maketitle

\begin{abstract}
Multi-phonon Raman scattering in semiconductor nanocrystals is treated
taking into account both adiabatic and non-adiabatic phonon-assisted optical
transitions. Because phonons of various symmetries are involved in
scattering processes, there is a considerable enhancement of intensities of
multi-phonon peaks in nanocrystal Raman spectra. Cases of strong and weak
band mixing are considered in detail. In the first case, fundamental
scattering takes place via internal electron-hole states and is participated
by $s$- and $d$-phonons, while in the second case, when the intensity of the
one-phonon Raman peak is strongly influenced by the interaction of an
electron and of a hole with interface imperfections (e.~g., with trapped
charge), $p$-phonons are most active. Calculations of Raman scattering
spectra for CdSe and PbS nanocrystals give a good quantitative agreement
with recent experimental results.
\end{abstract}


\section{Introduction}

The symmetry-breaking coupling of vibrational modes to degenerate electronic
states described first by Jahn and Teller \cite{JT37} has been manifested in
a wide range of transport and optical phenomena in various
electron-vibrational systems. The role of the Jahn-Teller effect in the
optical transitions of semiconductor nanocrystals has been recently studied
in Refs.\cite{Fomin,Fomin2} for the strong confinement regime, when the
radius of a nanocrystal $R$ is smaller than the Bohr radius of an exciton in
the bulk $R_{ex}$, and in Refs.\cite{Fedorov1,Fedorov} for the weak
confinement regime, when $R>R_{ex}$. In the case of a strong confinement,
when {\it all} the states of an exciton, including the ground state, are
degenerate due to band mixing, the Jahn-Teller effect substantially
influences the phonon-assisted optical transitions already in the framework
of the dipole approximation \cite{Fomin,Fomin2}. As a result, the optical
spectra of nanocrystals with $R<R_{ex}$ can be different from the
Franck-Condon progressions typical of the adiabatic systems.

Recently, multi-phonon peaks in Raman scattering spectra have been observed
in CdSe, CdSe$_{x}$S$_{1-x}$ and PbS nanocrystals (see, for example, Refs. %
\cite{Klein,Scamarcio,Krauss2}). Previously, the theoretical analysis of
multi-phonon Raman scattering in nanocrystals has been generally based on
the adiabatic approximation due to the seminal works of Huang and Rhys \cite%
{HR} and Pekar \cite{Pekar}. However, even taking into account band mixing,
the values of the Huang-Rhys factor for the nanocrystals of radius $R=1$ nm
to $R=5$ nm appear to be considerably smaller than unity. This result is at
variance with those derived from the experiments cited above. In the present
paper, the theoretical treatment is performed involving both adiabatic and
non-adiabatic optical transitions. The interaction of phonons with an
exciton in a degenerate state leads to {\em internal non-adiabaticity} of
nanocrystals (the Jahn-Teller effect), whilst the existence of exciton
levels separated by an energy comparable with that of the optical phonon
results in {\em external non-adiabaticity} (the pseudo Jahn-Teller effect).
The structural imperfections of the real interface between a nanocrystal and
a host medium can also substantially influence the probabilities of
phonon-assisted transitions, see e. g. Ref. \cite{Nirmal}. Moreover, these
imperfections (e.~g., trapped charge, localized at the interface, as
supposed in Refs. \cite{Krauss2,PRL79}) can increase the contribution of
non-adiabatic transitions into the scattering probability, as shown below.

In nanostructures, phonon spectra drastically differ from those in bulk
materials \cite{Klein,Englman,Richter,Roca,Chamberlain2}. Previously,
several models of polar optical phonons and of the electron-phonon
interaction have been developed for these structures. The {\em dielectric
continuum} model which is proposed in Ref. \onlinecite{Fuchs} and used for
planar \cite{Wendler,Lassing,Sawaki,Mori,ESR91}, cylindrical and spherical %
\cite{Klein,RE68} structures, exploits only electrostatic boundary
conditions. The alternative {\em hydrodynamic} model (see Refs. %
\onlinecite{Babiker,Trallerot,Ridley,Guillemot}) treats only mechanical
boundary conditions. Distinctions between the two models consist in a choice
of a complete orthogonal basis for the relative ionic displacement ${\bf u}%
\left( {\bf r}\right) $. It has been shown in Refs. \cite{Nash,Rucker} that
physical quantities expressed as a sum over all phonon modes (scattering
rates, polaron parameters etc.) slightly depend on a concrete basis. On the
other hand, calculated Raman spectra are sensitive to mechanical boundary
conditions. Various improvements of the dielectric continuum model \cite%
{HZ,Enderlein} are developed in order to obtain better agreement between
macroscopic and microscopic \cite{Rucker,HZ2} description of lattice
vibrations and with experimental data on Raman scattering \cite%
{Sood,SM89,Shields}. Within the continuum model of optical phonons in
spatially confined systems (Refs. %
\onlinecite{Roca,Chamberlain2,Chamberlain,Constantinou,Trallero}), both
electrostatic and mechanical boundary conditions are imposed on the ionic
displacement. The {\em multimode dielectric model} developed by us (see
Refs. \cite{PSS95,JLum}), in addition, explicitly takes into account a
finite number of phonon degrees of freedom in a quantum dot.

In Section 2, the set of phonon modes is determined using an effective
dielectric function of a nanocrystal in a finite-dimensional basis for the
field of the ionic displacements. In Section 3, multi-phonon resonant Raman
scattering in spherical nanocrystals is considered using the obtained
Hamiltonian of the electron-phonon interaction. The obtained analytical and
numerical results are discussed and compared with experimental data in
Section 4.

\section{Optical phonons in nanocrystals}

In the present investigation, longitudinal optical (LO) vibrations in
spherical quantum dots are considered on the basis of the multimode
dielectric model (see Refs. \cite{PSS95,JLum}). Polar optical phonons in a
semiconductor nanocrystal of a radius $R$ embedded into a polar medium can
be described within the continuum approach (when $R\gg a_{0},$ where $a_{0}$%
\ is the lattice constant) by the renormalized relative ionic displacement
vector 
\begin{equation}
{\bf w}_{k}\left( {\bf r}\right) =\frac{e_{k}^{\ast }{\bf u}_{k}\left( {\bf r%
}\right) }{V_{0k}\omega _{k,{\rm TO}}\sqrt{\varepsilon _{0}\left[
\varepsilon _{k}\left( 0\right) -\varepsilon _{k}\left( \infty \right) %
\right] }},  \label{W}
\end{equation}%
where $k=1$ for the nanocrystal, $k=2$ for the host medium. For the $k$-th
medium, ${\bf u}_{k}\left( {\bf r}\right) $ is the relative ionic
displacement, $e_{k}^{\ast }$ is the effective charge of an ion, $\omega _{k,%
{\rm TO}}$ is the frequency of the transverse optical phonon in the
Brillouin zone center, $V_{0k}$ is the volume of the elementary cell, $%
\varepsilon _{k}(\infty )$ and $\varepsilon _{k}(0)$ are the high-frequency
and static dielectric constants, respectively; $\varepsilon _{0}$ is the
permittivity of vacuum. Dynamics of the non-dispersive polar optical phonons
is determined by the Born-Huang equation \cite{Born} combined with Maxwell
equations. In order to analyze the time evolution of ${\bf w}_{k}\left( {\bf %
r}\right) $ taking into consideration the spatial dispersion, the
generalized equation of motion \cite{PSS95} is used: 
\begin{eqnarray}
\left( \omega _{k,{\rm TO}}^{2}-\omega ^{2}\right) {\bf w}_{k}\left( {\bf r}%
,\omega \right) &=&\omega _{k,{\rm TO}}\sqrt{\varepsilon _{0}\left[
\varepsilon _{k}\left( 0\right) -\varepsilon _{k}\left( \infty \right) %
\right] }{\bf E}_{k}\left( {\bf r},\omega \right)  \nonumber \\
&&+\int\limits_{V_{k}}{\bf T}_{k}\left( {\bf r}-{\bf x}\right) {\bf w}%
_{k}\left( {\bf x},\omega \right) \,d{\bf x}.  \label{EqMot}
\end{eqnarray}%
The right-hand side of Eq. (\ref{EqMot}) consists of two terms: (i) the
long-range force expressed by means of the macroscopic electric field ${\bf E%
}_{k}\left( {\bf r},\omega \right) $ induced by the ionic displacement
vector ${\bf w}_{k}\left( {\bf r},\omega \right) $; (ii) the dispersion term
of the short-range force with the spatial dispersion tensor ${\bf T}%
_{k}\left( {\bf r}\right) $. The action range of this force can be estimated
as a few lattice constants.

The set of equations to determine the polar optical phonon dynamics in both
the nanocrystal and the host medium consists of the equations (\ref{EqMot}),
the electrostatic Maxwell equations, and the equation relating the electric
field and the polarization to the electrostatic displacement for each
medium. The form of the spatial dispersion tensor ${\bf T}_{k}\left( {\bf r}%
\right) $\ is chosen in such a way that the bulk LO (TO) phonon dispersion $%
\omega _{{\rm LO(TO)}}\left( {\bf q}\right) $\ be reproduced\ when the
equation of motion (\ref{EqMot}) is solved in bulk. Further on, anisotropy
of the dispersion of bulk phonons is neglected: the bulk phonon frequencies
depend only on the wave number $q$.

For a nanocrystal, the aforesaid set of equations must be completed by
electrostatic and mechanical boundary conditions. In the present work, we
choose the following mechanical boundary conditions \cite{Constantinou,PSS95}
\begin{equation}
w_{k}^{\perp }=0\quad \mbox{at the
interfaces,}\quad k=1,2.  \label{ABC}
\end{equation}%
They lead to a double hybridization of LO and interface modes. For planar
multilayer structures, phonon spectra obtained in Ref. %
\onlinecite{Constantinou} when applying Eq. (\ref{ABC}) are in excellent
agreement with experimental Raman scattering data \cite{Haines}.

Owing to the spherical symmetry of a quantum dot, the eigenvectors of phonon
modes ${\bf g}_{lm}^{J}\left( {\bf r},\omega \right) $ correspond to a
definite phonon angular momentum $l$\ and its $z$-projection $m$. The index $%
J$\ labels LO and TO phonon branches. We seek these eigenvectors as a
superposition of orthogonal vector functions ${\bf f}_{lms}^{J}\left( {\bf r}%
\right) $\ which satisfy Eq. (\ref{ABC}):%
\begin{equation}
{\bf g}_{lm}^{J}\left( {\bf r},\omega \right) =\sum_{s}U_{l,s}^{J}\left(
\omega \right) {\bf f}_{lms}^{J}\left( {\bf r}\right) .  \label{Expansion}
\end{equation}%
This basis is subdivided into three subsets describing:

\noindent (i) LO phonons with definite values of the angular momentum $l$,
of its projection $m$, and of the radial quantum number $s$, 
\[
{\bf f}_{lms}^{{\rm LO}}\left( {\bf r}\right) =\frac{\sqrt{2/R}}{j_{l}\left(
b_{l,s}\right) \sqrt{b_{l,s}^{2}-l\left( l+1\right) }}\left\{ \frac{b_{l,s}}{%
R}j_{l}^{\prime }\left( \frac{b_{l,s}r}{R}\right) ^{{}}{\bf Y}_{lm}^{\left(
-1\right) }\left( {\bf n}\right) \phantom{\frac{\sqrt{l}}{r}}\right. 
\]%
\begin{equation}
\left. +\frac{\sqrt{l\left( l+1\right) }}{r}j_{l}\left( \frac{b_{l,s}r}{R}%
\right) {\bf Y}_{lm}^{\left( 1\right) }\left( {\bf n}\right) \right\} ,
\label{LObasis}
\end{equation}%
where ${\bf n}={\bf r}/r$, $j_{l}\left( x\right) $ is a spherical Bessel
function, $b_{l,s}$ is the $s$-th zero of its derivative $j_{l}^{\prime
}\left( x\right) $, ${\bf Y}_{lm}^{\left( \lambda \right) }\left( {\bf n}%
\right) $ ($\lambda =0,\pm 1$) is a spherical vector of a definite parity;

\noindent (ii) TO phonons of the ``electric'' type, 
\[
{\bf f}_{lms}^{{\rm TO},E}\left( {\bf r}\right) =\frac{\sqrt{2/R}}{%
a_{l,s}j_{l+1}\left( a_{l,s}\right) }\left\{ \frac{\sqrt{l\left( l+1\right) }%
}{r}j_{l}\left( \frac{a_{l,s}r}{R}\right) {\bf Y}_{lm}^{\left( -1\right)
}\left( {\bf n}\right) \right. 
\]%
\begin{equation}
\left. \phantom{\frac{\sqrt{l}}{r}}+\frac{a_{l,s}}{R}\left[ j_{l}^{\prime
}\left( \frac{a_{l,s}r}{R}\right) +\frac{R}{a_{l,s}r}j_{l}\left( \frac{%
b_{l,s}r}{R}\right) \right] {\bf Y}_{lm}^{\left( 1\right) }\left( {\bf n}%
\right) \right\} ,  \label{TOEbasis}
\end{equation}%
where $a_{l,s}$ is the $s$-th zero of $j_{l}\left( x\right) $;

\noindent (iii) TO phonons of the ``magnetic'' type, 
\begin{equation}
{\bf f}_{lms}^{{\rm TO},M}\left( {\bf r}\right) =\frac{\sqrt{2}}{%
R^{3/2}j_{l+1}\left( a_{l,s}\right) }j_{l}\left( \frac{a_{l,s}r}{R}\right) 
{\bf Y}_{lm}^{\left( 0\right) }\left( {\bf n}\right) .  \label{TOMbasis}
\end{equation}%
Dimensionalities of the subsets (\ref{LObasis})-(\ref{TOMbasis}) are
implicitly determined by the inequalities: 
\begin{equation}
\begin{array}{c}
b_{l,s}\leq \pi R/a_{0},\qquad \mbox{for LO modes}; \\ 
a_{l,s}\leq \pi R/a_{0},\qquad \mbox{for TO modes},%
\end{array}
\label{Dimension}
\end{equation}%
which express the fact that the ``wavelength'' of an optical phonon cannot
be smaller than the double lattice constant.

Using the expansion (\ref{Expansion}) for a joint solution of Eq. (\ref%
{EqMot}) and of the Maxwell equations with electrostatic boundary
conditions, we arrive at the dispersion equation 
\begin{equation}
\frac{l+1}{\varepsilon _{1}\left( l,\omega \right) }+\frac{l}{\varepsilon
_{2}\left( \omega \right) }=0,  \label{DispEq}
\end{equation}%
where $\varepsilon _{2}\left( \omega \right) $ is the dielectric function of
the host medium. The function $\varepsilon _{1}\left( l,\omega \right) $ is
determined by the formula 
\begin{equation}
\frac{1}{\varepsilon _{1}\left( l,\omega \right) }\equiv \frac{1}{%
\varepsilon _{1}\left( \infty \right) }\left( 1-\sum_{s}\chi _{l,s}\frac{%
\omega _{1,{\rm LO}}^{2}\left( 0\right) -\omega _{1,{\rm TO}}^{2}\left(
0\right) }{\omega _{1,{\rm LO}}^{2}\left( Q_{l,s}\right) -\omega ^{2}}%
\right) ,  \label{EffEps}
\end{equation}%
where $Q_{l,s}\equiv b_{l,s}/R,$ and the coefficients $\chi _{l,s}$ are: 
\begin{equation}
\chi _{l,s}=\frac{2l}{b_{l,s}^{2}-l\left( l+1\right) }.  \label{Oscill}
\end{equation}%
These coefficients possess the property $\sum_{s=1}^{\infty }\chi _{l,s}=1.$

We can interpret $\varepsilon _{1}\left( l,\omega \right) $\ as the
effective dielectric function of the quantum dot, using a formal analogy of
the dispersion equation (\ref{DispEq}) with that of the dielectric continuum
model (cf. Eq. (13) of Ref. \cite{Klein}). In the ``non-dispersive'' limit,
when we set in Eq. (\ref{EffEps}) $\omega _{1,{\rm LO}}\left( Q_{l,s}\right)
=\omega _{1,{\rm LO}}\left( 0\right) ,$\ the function $\varepsilon
_{1}\left( l,\omega \right) $\ becomes equal to the ``bulk'' dielectric
function $\varepsilon _{1}\left( \omega \right) ,$\ and Eq. (\ref{DispEq})
turns into the dispersion equation of the dielectric continuum model \cite%
{Klein}. The latter equation provides, in particular, interface phonon
frequencies $\omega _{j,{\rm I}}\left( l\right) $\ [$j=1,2;$\ $\omega _{1,%
{\rm TO}}\left( 0\right) <\omega _{j,{\rm I}}\left( l\right) <\omega _{1,%
{\rm LO}}\left( 0\right) $].

Basis vectors of phonon modes are denoted as ${\bf g}_{lm\eta }^{J}\left( 
{\bf r}\right) $, where the index $\eta $ labels the roots of Eq. (\ref%
{DispEq}) $\omega _{l,\eta }$ at a definite value of the angular momentum $l$%
. Generally speaking, LO phonon modes cannot be subdivided into bulk-like
and interface ones. Therefore, they can be considered as hybrids of both
these types. The following explicit expression is obtained for the
coefficients in Eq. (\ref{Expansion}): 
\begin{equation}
U_{l,s}^{{\rm LO}}\left( \omega _{l,\eta }\right) =\frac{C_{l}\left( \omega
_{l,\eta }\right) }{\sqrt{b_{l,s}^{2}-l\left( l+1\right) }\left[ \omega _{1,%
{\rm LO}}^{2}\left( Q_{l,s}\right) -\omega _{l,\eta }^{2}\right] }.
\label{ExpCoef}
\end{equation}%
The transformation (\ref{Expansion}) is unitary (see Ref. \cite{PSS95}). The
normalization constant $C_{l}\left( \omega _{l,\eta }\right) $\ is then
found to be%
\begin{equation}
C_{l}\left( \omega _{l,\eta }\right) =\left[ \sum_{s}\frac{1}{\left[
b_{l,s}^{2}-l\left( l+1\right) \right] \left[ \omega _{1,{\rm LO}}^{2}\left(
Q_{l,s}\right) -\omega _{l,\eta }^{2}\right] ^{2}}\right] ^{-1/2}.
\label{Norma}
\end{equation}%
Finally, the phonon Hamiltonian takes the form: 
\begin{equation}
\hat{H}_{L}=\sum_{\nu }\hbar \omega _{\nu }\left( \hat{a}_{\nu }^{\dagger }%
\hat{a}_{\nu }+\frac{1}{2}\right) ,  \label{HL}
\end{equation}%
while the Hamiltonian of the electron-phonon interaction is 
\begin{equation}
\hat{H}_{e-L}=\sum_{\nu }\left( \hat{\gamma}_{\nu }\hat{a}_{\nu }+\hat{\gamma%
}_{\nu }^{\dagger }\hat{a}_{\nu }^{\dagger }\right) ,  \label{Hint}
\end{equation}%
where the set of indexes $\nu =\left( l,m,\eta \right) $ labels the obtained
LO modes. The amplitudes of the electron-phonon interaction can be
explicitly expressed in terms of the unitary transformation coefficients\ as
follows: 
\begin{eqnarray}
\gamma _{lm\eta }\left( {\bf r}\right) &=&\frac{2\hbar \omega _{1,{\rm LO}}}{%
\sqrt{R}}\left( \frac{\hbar }{m\omega _{1,{\rm LO}}}\right) ^{1/2}\left( 
\sqrt{2}\pi \alpha _{1}\right) ^{1/2}{\rm Y}_{lm}\left( {\bf n}\right) 
\nonumber \\
&&\times \sum_{s}\frac{U_{l,s}\left( \omega _{l,\eta }\right) }{\sqrt{%
b_{l,s}^{2}-l\left( l+1\right) }}\left[ \frac{j_{l}\left( b_{l,s}r/R\right) 
}{j_{l}\left( b_{l,s}\right) }-\frac{\left( l+1\right) \varepsilon
_{2}\left( \omega _{l,\eta }\right) }{l\varepsilon _{1}+\left( l+1\right)
\varepsilon _{2}\left( \omega _{l,\eta }\right) }\left( \frac{r}{R}\right)
^{l}\right] ,  \label{EPI1}
\end{eqnarray}%
where $\alpha _{1}$ is the Fr\"{o}hlich electron-phonon coupling constant of
a nanocrystal.\ The Hamiltonian of the exciton-phonon interaction has the
form similar to Eq. (\ref{Hint}) with the replacement of $\hat{\gamma}_{\nu
} $\ by the exciton-phonon interaction amplitudes 
\begin{equation}
\hat{\beta}_{\nu }\left( {\bf r}_{e},{\bf r}_{h}\right) \equiv \hat{\gamma}%
_{\nu }\left( {\bf r}_{e}\right) -\hat{\gamma}_{\nu }\left( {\bf r}%
_{h}\right) .  \label{beta}
\end{equation}

It is worth noting that in CdSe and in PbS, where dispersion of bulk LO
phonons is strong (see Refs. \cite{SM89,Elcombe,Krauss1}), the LO phonon
modes as derived from Eq. (\ref{DispEq}) appear to be hybrids of bulk-like
and interface vibrations.

\section{Raman spectra}

Within the context of the long-wavelength approximation, the interaction of
an electron with an electromagnetic field is described by the operator $\hat{%
V}\left( t\right) =\hat{V}_{I}e^{-i\Omega _{I}t}+\hat{V}_{S}^{\dagger
}e^{i\Omega _{S}t}$, where the terms $\hat{V}_{I}$ and $\hat{V}_{S}^{\dagger
}$ correspond, respectively, to the absorption of a photon with the
frequency $\Omega _{I}$ (incoming light) and to the emission of a photon
with the frequency $\Omega _{S}$ (scattered light). The interaction
amplitude $\hat{V}_{I(S)}$ is proportional to the projection of the electron
dipole momentum operator ${\bf \hat{d}}$ on the polarization vector ${\bf e}%
^{I(S)}$ of the relevant wave: $\hat{d}^{I(S)}={\bf e}^{I(S)}\cdot {\bf \hat{%
d}}$. From the second-order perturbation theory, the transition probability
between an initial $\left| i\right\rangle $ and a final $\left|
f\right\rangle $ state is 
\begin{equation}
w_{i\rightarrow f}=\frac{2\pi }{\hbar ^{4}}\left| \sum\limits_{m}\frac{%
\left\langle f\left| \hat{V}_{S}^{\dagger }\right| m\right\rangle
\left\langle m\left| \hat{V}_{I}\right| i\right\rangle }{\omega _{mi}-\Omega
_{I}+i\delta }\right| ^{2}\delta \left( \omega _{fi}-\Omega _{I}+\Omega
_{S}\right) ;\quad \delta \rightarrow +0.  \label{Probab}
\end{equation}%
Here, $\omega _{mi}$ and $\omega _{fi}$ are transition frequencies. For
phonon-assisted Raman scattering in a semiconductor nanocrystal, $\left|
i\right\rangle $, $\left| f\right\rangle $, and $\left| m\right\rangle $ are
quantum states of the exciton-phonon system. Both the initial and the final
states contain no charge carriers (electrons or holes), so that these states
are described by a direct product of a wave function of free phonons with
that of the exciton vacuum. Intermediate states $\left| m\right\rangle $ are
eigenstates of the Hamiltonian 
\begin{equation}
\hat{H}=\hat{H}_{ex}+\hat{H}_{ph}+\hat{H}_{ex-ph},
\end{equation}%
where $\hat{H}_{ex}$ is the Hamiltonian of an exciton, $\hat{H}_{ph}$ is the
phonon Hamiltonian (\ref{HL}), and $\hat{H}_{ex-ph}$ is the Hamiltonian of
the exciton-phonon interaction.

For definite polarizations of the incoming and the scattered light, the
scattering probability is obtained by averaging Eq. (\ref{Probab}) over the
initial states and by summing over the final ones. Since the phonon
Hamiltonian $\hat{H}_{L}$\ is quadratic and the Hamiltonian of the
exciton-phonon interaction $\hat{H}_{ex-L}$\ is linear in phonon creation
and annihilation operators, an analytical averaging over the equilibrium
phonon ensemble can be performed in Eq. (\ref{Probab}). As a result, we can
express the shape of the Raman spectrum in terms of time-ordered operators
averaged over exciton states only: 
\begin{eqnarray}
W\left( \Omega _{I},{\bf e}_{I},\Omega _{S},{\bf e}_{S}\right)
&=&\int\limits_{-\infty }^{\infty }dt\ e^{i\left( \Omega _{S}-\Omega
_{I}\right) t}\int\limits_{0}^{\infty }d\tau \int\limits_{0}^{\infty
}d\sigma \ e^{-\delta \left( \tau +\sigma \right) -i\Omega _{I}\left( \tau
-\sigma \right) }  \nonumber \\
&&\times \sum_{\mu _{1}\mu _{2}}\sum_{\mu _{1}^{\prime }\mu _{2}^{\prime
}}\left( d_{\mu _{1}^{\prime }}^{I}d_{\mu _{2}}^{S}\right) ^{\ast }d_{\mu
_{1}}^{I}d_{\mu _{2}^{\prime }}^{S}e^{i\left( \tilde{\omega}_{\mu _{1}}\tau -%
\tilde{\omega}_{\mu _{1}^{\prime }}\sigma \right) }  \nonumber \\
&&\times \left\langle \mu _{1}^{\prime }\left| \left\langle \mu _{2}\left| 
{\rm T}_{s^{\prime }}{\rm T}_{s}^{-}\exp \left\{ \Phi \left[ \hat{\beta},%
\hat{\beta}^{\prime }\right] \right\} \right| \mu _{1}\right\rangle \right|
\mu _{2}^{\prime }\right\rangle .  \label{Formfunc}
\end{eqnarray}%
Here, $\tilde{\omega}_{\mu }$\ are the frequencies and $d_{\mu
}^{I(S)}\equiv \langle \mu |\hat{d}^{I(S)}|0\rangle $\ are the dipole matrix
elements for a transition from the exciton vacuum $|0\rangle $\ to the
eigenstates $|\mu \rangle $\ of the Hamiltonian $\hat{H}_{ex}.$ The
``influence phase'' of the phonon subsystem $\Phi \left[ \hat{\beta},\hat{%
\beta}^{\prime }\right] $\ is the following operator: 
\begin{eqnarray}
\Phi \left[ \hat{\beta},\hat{\beta}^{\prime }\right] &=&\frac{1}{\hbar ^{2}}%
\sum_{\nu }\left\{ \int\limits_{0}^{\tau }ds\int\limits_{0}^{\sigma
}ds^{\prime }T_{\omega _{\nu }}^{\ast }\left( t+s-s^{\prime }\right) \hat{%
\beta}_{\nu }^{\prime }\left( s^{\prime }\right) \hat{\beta}_{\nu }^{\dagger
}\left( s\right) \right.  \nonumber \\
&&-\left. \int\limits_{0}^{\tau }ds\int\limits_{0}^{s}ds^{\prime }T_{\omega
_{\nu }}\left( s-s^{\prime }\right) \hat{\beta}_{\nu }\left( s\right) \hat{%
\beta}_{\nu }^{\dagger }\left( s^{\prime }\right) -\int\limits_{0}^{\sigma
}ds\int\limits_{0}^{s}ds^{\prime }T_{\omega _{\nu }}^{\ast }\left(
s-s^{\prime }\right) \hat{\beta}_{\nu }^{\prime }\left( s^{\prime }\right) 
\hat{\beta}_{\nu }^{\dagger }\left( s\right) \right\} .  \label{Phase}
\end{eqnarray}%
In Eqs. (\ref{Formfunc}) and (\ref{Phase}), primed (non-primed)
exciton-phonon amplitudes are (anti)chronologically ordered operators, which
act, respectively, on primed (non-primed) exciton states. The phonon Green's
function 
\begin{equation}
T_{\omega }\left( t\right) =\frac{\cosh \left[ \omega \left( t-\hbar
/2k_{B}T\right) \right] }{\sinh \left( \hbar \omega /2k_{B}T\right) }
\label{Thornber}
\end{equation}%
describes the phonon emission and absorption processes.

Because electron-phonon coupling is weak in nanocrystals, such as CdSe, CdSe$%
_{x}$S$_{1-x}$ and PbS, the $K$-phonon scattering intensity, corresponding
to a definite combinatorial frequency $\sum\limits_{j=1}^{K}\omega _{\nu
_{j}}$, can be analyzed to the leading ($K$-th) order in the electron-phonon
coupling constant \cite{Hayes}. The scattering intensity within this {\em %
leading-term approach }is then expressed through a squared modulus of the
scattering amplitude: 
\begin{equation}
F_{K}^{\left( \pm \right) }\left( \nu _{1},...,\nu _{K}\right)
\,=\,\sum_{\mu \ _{0}...\mu _{K}}\frac{d_{\mu _{0}}^{I}\left( d_{\mu
_{K}}^{S}\right) ^{\ast }}{\tilde{\omega}_{\mu _{0}}-\Omega _{I}+i\tilde{%
\Gamma}_{\mu _{0}}}\prod_{j=1}^{K}\frac{\left\langle \mu _{j}\left| \hat{%
\beta}_{\nu _{j}}\right| \mu _{j-1}\right\rangle }{\tilde{\omega}_{\mu
_{j}}-\Omega _{I}\pm \sum\limits_{k=1}^{j}\left( \omega _{\nu _{k}}\pm
i\Gamma _{\nu _{k}}\right) +i\tilde{\Gamma}_{\mu _{j}}},  \label{F}
\end{equation}%
where $\tilde{\Gamma}_{\mu }$ is the inverse lifetime of an exciton in the
state $\left| \mu \right\rangle $, while $\Gamma _{\nu }$ is the inverse
lifetime of a phonon of the mode $\nu $.

The following treatment is related to the regime of strong confinement $%
R<R_{ex}$, where $R_{ex}$\ is the exciton Bohr radius. In this case, the
Coulomb electron-hole interaction can be treated as a perturbation, and the
concept of the exciton means the same as that of the electron-hole pair.

If one considers states of the electron-hole pair $\left| \mu
_{j}\right\rangle $ in Eq. (\ref{F}) within the model of simple bands, then
diagonal matrix elements of amplitudes of the electron-phonon interaction on
the wave function of the ground electron-hole state vanish, what suppresses
the one-phonon Raman scattering. This suppression can be removed by one of
the following mechanisms: (i) the Coulomb electron-hole interaction \cite%
{Rodriguez}, (ii) band mixing \cite{X89}, (iii) the influence of interface
imperfections on the exciton wave functions \cite{Nirmal}, (iv)
non-adiabatic transitions with the participation of virtual phonons. The
latter mechanism exists only beyond the leading-term approach.

Two particular cases are of interest when one of these mechanisms dominates:
(i) strong band mixing (CdSe in glass \cite{Klein,Scamarcio}), (ii) weak
band mixing (PbS in PVA \cite{Krauss2}) when one-phonon transitions are
determined mainly by the scattering of an electron and of a hole by the
potential due to the imperfections. It is worth recalling, that in the
strong-confinement regime, the Coulomb interaction only slightly influences
the electron-hole states and the Raman spectra.

For a CdSe nanocrystal, exciton states are analyzed here using the spherical
model of the exciton Hamiltonian \cite{Baldereschi} supplemented by terms
describing the electron-hole exchange interaction \cite{Nirmal2}: 
\begin{eqnarray}
\hat{H}_{ex} &=&\frac{1}{2m_{e}}{\bf \hat{p}}_{e}^{2}+\frac{\gamma _{1}}{%
2m_{0}}{\bf \hat{p}}_{h}^{2}-\frac{\gamma _{2}}{9m_{0}}\left( {\bf \hat{P}}%
^{\left( 2\right) }\cdot {\bf \hat{J}}^{(2)}\right)  \nonumber \\
&&+V_{C}\left( {\bf r}_{e},{\bf r}_{h}\right) -\frac{2}{3}\epsilon
_{exch}a_{0}^{3}\delta \left( {\bf r}_{e}-{\bf r}_{h}\right) \left( \widehat{%
\vec{\sigma}}\cdot {\bf \hat{J}}\right) ,  \label{Ham}
\end{eqnarray}%
where ${\bf \hat{p}}_{e}$ and ${\bf \hat{p}}_{h}$ (${\bf r}_{e}$ and ${\bf r}%
_{h}$) are the electron and hole momenta (coordinate vectors); $\gamma _{1}$%
\ and $\gamma _{2}$\ are the Luttinger parameters, $m_{0}$\ and $m_{e}$\ are
the bare electron mass and the electron band mass; ${\bf \hat{P}}^{(2)}$ and 
${\bf \hat{J}}^{(2)}$ are irreducible second-rank tensors of the momentum
and of the spin-$\frac{3}{2}$\ angular momentum of a hole, respectively. The
potential of the Coulomb attraction between an electron and a hole in a
spherical nanocrystal $V_{C}\left( {\bf r}_{e},{\bf r}_{h}\right) $\ is used
from Ref. \onlinecite{PSS95}. In Eq. (\ref{Ham}), the last term, which is
proportional to the scalar product of the electron ($\widehat{\vec{\sigma}}$%
) and hole (${\bf \hat{J}}$) spin operators, describes the electron-hole
exchange interaction\ characterized by the strength constant $\epsilon
_{exch}$. In CdSe, this constant is equal to 320 meV \cite{Nirmal2}. The
Coulomb and the exchange interactions between an electron and a hole are
treated as perturbations when determining the exciton states. To the zeroth
order in these perturbations, the exciton states $\left| \mu \right\rangle $%
\ are characterized by a definite electron spin projection $\sigma $\ and a
total angular momentum of a hole $F$\ with the $z$-projection $M$: 
\begin{equation}
\left| \mu \right\rangle \equiv \left| 1S,\sigma ;{\sf K},M\right\rangle
=\Psi _{1S}^{e}\left( \sigma \right) \Phi _{{\sf K}}^{h}\left( M\right) ,
\label{WF}
\end{equation}%
where ${\sf K}=n$S$_{3/2}$, $n$P$_{1/2}$, $n$P$_{3/2}$, $n$P$_{5/2}$,\ etc.
The index $n$\ labels the solutions of the equations for the radial
components of the hole wave function \cite{X89}.

In contrast to CdSe, the conduction and valence bands in PbS \cite{Lin} are
both non-degenerate. As a result, band mixing in PbS nanocrystals exerts a
very small influence on the matrix elements of the exciton-phonon
interaction. Note that the exciton-phonon coupling due to this mixing in PbS
quantum dots calculated using the four-band envelope-function formalism \cite%
{Krauss2} is over 2 orders of magnitude smaller than the value of CdSe
quantum dots at all radii. Hence, the influence of boundary structure
imperfections on the probabilities of one-phonon optical transitions becomes
of paramount importance. This gives us a reason to consider electron-hole
states in PbS within the model of simple bands. The influence of interface
imperfections is modeled by the potential expanded in spherical harmonics ${Y%
}_{lm}\left( \vartheta ,\varphi \right) $: 
\begin{equation}
U_{{\rm imp}}\left( {\bf r}\right) =\sum_{l=1}^{\infty
}\sum_{m=-l}^{l}U_{lm}\left( \frac{r}{R}\right) ^{l}{Y}_{lm}\left( \vartheta
,\varphi \right) ,  \label{Random}
\end{equation}%
which obeys the Laplace equation.\ Exciton quantum states $\left| \mu
_{j}\right\rangle $ in Eq. (\ref{F}) are calculated in the first
perturbation approximation on the potential $U_{{\rm imp}}\left( {\bf r}%
\right) $. Calculated scattering intensities are then averaged on the
Gaussian distribution of random amplitudes $U_{lm}$ with the variance $U_{0}$%
\ as a fitting parameter. Eq. (\ref{Random}) describes, in particular, the
electrostatic potential, induced by {\em trapped charge}. This charge can
build up on the nanocrystals during the steady-state Raman measurements \cite%
{Krauss2,PRL79}.

In Raman scattering experiments \cite{Klein,Scamarcio,Krauss2}, the
frequency of the incoming light is chosen to be in resonance with the
exciton ground-state energy. Hence, only several lowest levels should be
considered when calculating scattering amplitudes (\ref{F}). Two types of 
{\it selection rules} manifest themselves in the phonon-assisted Raman
scattering. The first group of selection rules results from the symmetry
properties of the {\it dipole matrix elements} $d_{\mu }^{I}$ and $d_{\mu
^{\prime }}^{S}$: (i) for parallel polarizations (${\bf e}^{I}\parallel {\bf %
e}^{S}$), the projection of the exciton angular momentum is preserved in the
scattering process, $M=M^{\prime }$; (ii) for crossed polarizations (${\bf e}%
^{I}\perp {\bf e}^{S}$), $M=M^{\prime }\pm 1$.

The second group of selection rules is determined by the symmetry properties
of matrix elements of the {\it exciton-phonon interaction} amplitudes $%
\left\langle \mu _{j}\left| \hat{\beta}_{\nu _{j}}\right| \mu
_{j-1}\right\rangle $. Only the following phonons take part in the
one-phonon scattering: (i) phonons with $m=0$, in parallel polarizations;
(ii) phonons with $m=\pm 1$, in crossed polarizations. The selection rules
for the one-phonon Raman scattering in spherical nanocrystals were analyzed
in Ref. \cite{Chamberlain2} supposing a non-degenerate parabolic band
structure, and the conclusion drawn was that only $s$-phonons (with $l=0$)
are active in the dipole scattering processes. This conclusion cannot be
applied to nanocrystals with a degenerate valence band, where due to band
mixing $s$-phonons and $d$-phonons (with $l=2)$ appear to be active in the
one-phonon Raman scattering.

In the multi-phonon scattering amplitudes of Eq.\thinspace (\ref{F}) with
the phonon number $K\geq 2$, each matrix element of the exciton-phonon
interaction contains at least one {\it intermediate} quantum state, which
can be optically active or inactive. Since in the scattering amplitudes of
Eq.\thinspace (\ref{F}), the summation is carried out over all intermediate
states, phonons of arbitrary angular momenta can participate in multi-phonon
transitions assisted by more than one phonon. It should be mentioned, that
the adiabatic approximation allows participation of only $s$-phonons in
multi-phonon scattering processes \cite{Klein}.

It is worth noting that the one-phonon Raman scattering assisted by $p$%
-phonons is {\em forbidden} in the case when a nanocrystal has the inversion
symmetry. However, the potential due to imperfections [Eq. (\ref{Random})]
contains both even and odd terms and hence {\em breaks this symmetry down}.
Owing to the breakdown of the inversion symmetry, the $p$-phonons (and other
phonons of the odd parity) can be active in the one-phonon Raman scattering.

\section{Discussion of results}

For the calculation of Raman spectra for CdSe quantum dots, the following
values of parameters are used: $\varepsilon _{1}(\infty )=6.23$, $%
\varepsilon _{1}(0)=9.56$\ (Ref. \cite{Nomura}), the electron band mass $%
m_{e}=0.11m_{0},$\ the bulk energy band gap $E_{g}=1.9$\ eV (Ref. \cite%
{Landolt}), the Luttinger parameters $\gamma _{1}=2.04$, $\gamma _{2}=0.58$\
(Ref. \cite{Norris}), and $\varepsilon _{2}=2.25$\ (Ref. \cite{Klein}).

For PbS quantum dots, we have taken $E_{g}=0.307$ eV (Ref. \cite{MPZ64}), $%
\varepsilon _{1}\left( \infty \right) =18.5,$ $\varepsilon _{1}\left(
0\right) =190$ (Ref. \cite{Kartheuser}). The energy of the size quantization
in the PbS nanocrystals under consideration (with $R\sim 1$ to 2 nm) is
considerably larger than the bulk band gap. For such a large energy, the
dispersion law of both an electron and a hole is substantially
non-parabolic, so that we cannot use the effective-mass approach even for
simple bands. In this view, we determine exciton energies from the
experimental absorption spectra of Ref. \cite{Krauss2}. The peak positions
observed in Ref. \cite{Krauss2} for PbS nanocrystals of the radius 1.5 nm
correspond to the energies $E_{exc}(1s,1s)\approx 2.06$ eV, $E_{exc}\left(
1p,1p\right) \approx 3.14$ eV, $E_{exc}(1d,1d)\approx 4.18$ eV. Since the
conduction and valence bands in PbS are known to be approximately mirror
images of each other, the energies of transitions between the lowest states
of an electron and of a hole can be estimated as $E_{e(h)}\left( 1p\right)
-E_{e(h)}\left( 1s\right) \approx \left[ E_{exc}\left( 1p,1p\right)
-E_{exc}(1s,1s)\right] /2\approx 0.54$ eV, $E_{e(h)}\left( 1d\right)
-E_{e(h)}\left( 1s\right) \approx \left[ E_{exc}\left( 1d,1d\right)
-E_{exc}(1s,1s)\right] /2\approx 1.06$ eV.

The Raman scattering spectrum, calculated for a Gaussian ensemble of CdSe
quantum dots in glass and compared with experimental data of Ref. \cite%
{Klein}, is shown in Fig. 1. The relative size dispersion $\delta \equiv 
\sqrt{\left\langle \Delta R^{2}\right\rangle }/\left\langle R\right\rangle $
is taken 10\% (see Ref. \cite{Klein}). The parameter $\tilde{\Gamma}_{\mu
}\equiv \tilde{\Gamma}$ describing the linewidth of exciton states is taken $%
\tilde{\Gamma}=0.3\omega _{1,{\rm LO}}$, that is close to values measured in
Ref. \cite{Mittleman}.

The structure of the phonon-assisted Raman scattering peaks for the
spherical CdSe nanocrystals appears to be as follows. According to the
selection rules discussed above, only $s$- and $d$-phonons are active in the
one-phonon spectrum. In crossed polarizations, transitions assisted by $s$
-phonons are forbidden. Hence, the one-phonon spectrum is determined only by 
$d$-phonons with $m=\pm 1$. In CdSe nanocrystals, the probabilities of
transitions are determined mainly by the considerable effect of mixing
between bands of heavy and light holes, while the role of interface
imperfections is of minor importance. It is worth noting that, as a result
of {\it non-adiabatic transitions} within the state (1S,1S$_{3/2}$), the
contribution to scattering intensity due to $d$-phonons is significantly
enhanced compared to that calculated \cite{Nomura,Chen} in the adiabatic
approximation. This enhancement is a result of the Jahn-Teller effect. In
the multi-phonon scattering processes (with $K\geq 2$), the participation of
phonons of other types, in particular, $p$- and $f$-phonons, is permitted
under the above selection rules. It is also worth noting that transitions
assisted by $p$-phonons through excited states (1S,1P$_{3/2}$) and (1S,1P$%
_{5/2}$) bring a key contribution into the multi-phonon scattering
probabilities for spherical CdSe quantum dots with typical radii $R\sim 1$
to 4 nm. Recently, in Ref. \cite{Rodriguez}, the multi-phonon Raman
scattering in spherical semiconductor nanocrystals has been treated within
the model \cite{Roca} for the phonon spectrum and for the electron-phonon
interaction. In Ref. \cite{Rodriguez}, no imperfections and no non-adiabatic
effects are taken into account. Relative overtone intensities, calculated in
Ref. \cite{Rodriguez} using the effective-mass approximation for excitonic
states, quantitatively differ from the experimental data of Ref. \cite{Klein}%
. In order to provide a quantitative description of the Raman spectra, in
Ref. \cite{Rodriguez} a correction factor has been introduced. As seen from
Fig. 1, our results compare well with those of Ref. \cite{Klein} without
using any correction factors.

The Raman scattering spectrum obtained for an ensemble of PbS nanocrystals
with a relative size dispersion $\pm 4\%$ (Ref. \cite{Krauss2}) is shown in
Fig. 2. This spectrum is calculated with $U_{0}=0.036$ eV, which is much
less than the typical energy difference between exciton levels. Furthermore,
the value $\Gamma _{\nu }\equiv \Gamma =15$ cm$^{-1}$ corresponds to the
experimentally observed peak broadening which is attributed to a finite
phonon lifetime (see Ref. \cite{Krauss2}). The fundamental scattering
intensity in PbS nanocrystals due to both adiabatic and non-adiabatic
transitions is determined by trapped charge and is proportional to $%
U_{0}^{2} $, while overtone peaks are formed mainly as a result of
non-adiabatic transitions through the lowest excited states of an exciton.

The main contribution to both fundamental and overtone bands in PbS quantum
dots comes from the exciton-phonon interaction with $1p$- and $2p$-phonons.
This contrasts both with the results of the adiabatic approach, which
implies a domination of peaks corresponding to $s$-phonons, and with those
for CdSe quantum dots, where $s$- and $d$-phonons are active in the
fundamental scattering. Peak positions given by the adiabatic theory differ
significantly from those of the experimental data \cite{Krauss2}.

As a result of considerable LO phonon dispersion in bulk material, the
phonon frequencies in PbS quantum dots vary substantially for different
modes, i.~e. $\omega _{1s}=235$ cm$^{-1}$, $\omega _{1p}=217$ cm$^{-1}$, $%
\omega _{2p}=192$ cm$^{-1}$, $\omega _{1d}=229$ cm$^{-1}$, $\omega _{2d}=114$
cm$^{-1}$. Therefore, the phenomena contributing to Raman peaks can be
easily identified. In particular, main peaks in both fundamental and
overtone bands can be confidently assigned to $p$-phonons. It is evident
from Fig. 2, that a satisfactory agreement exists between the calculated
Raman spectrum and the experimental results \cite{Krauss2} with respect to
both ratios between spectral intensities and peak positions.

\acknowledgements

We are grateful to V. N. Gladilin for valuable discussions. This work has
been supported by GOA BOF UA 2000, I.U.A.P., F.W.O.-V. projects Nos.
G.0287.95, 9.0193.97, G.02774.01N and the W.O.G. WO.025.99N (Belgium).
E.P.P. acknowledges with gratitude kind hospitality during his visits to UIA.

\newpage

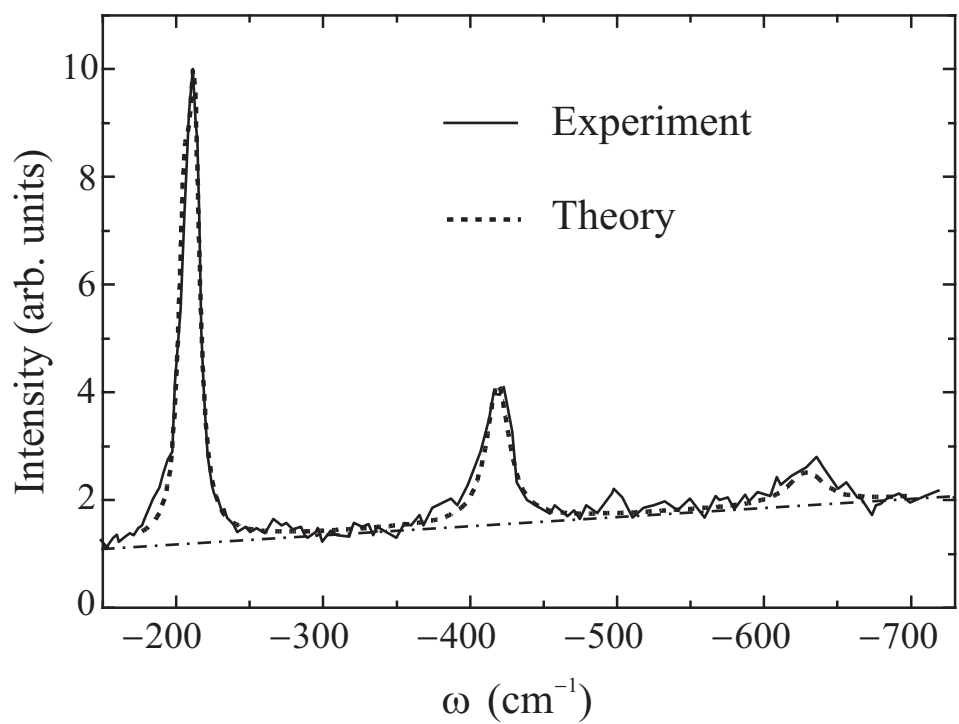
\noindent Fig. 1. Multi-phonon Raman scattering spectrum for the ensemble of
CdSe nanocrystals with $\left\langle R\right\rangle =2$ nm compared with the
experimental data of Ref. \cite{Klein}. The thin dashed line is the
luminescence background. The following parameter values are used: $%
\varepsilon _{1}(\infty )=6.23$, $\varepsilon _{1}(0)=9.56$ \cite{Nomura}, $%
m_{e}=0.11m_{0},$ $E_{g}=1.9$ eV \cite{Landolt}, $\gamma _{1}=2.04$, $\gamma
_{2}=0.58$ \cite{Norris}, and $\varepsilon _{2}=2.25$ (Ref. \cite{Klein}).

\bigskip

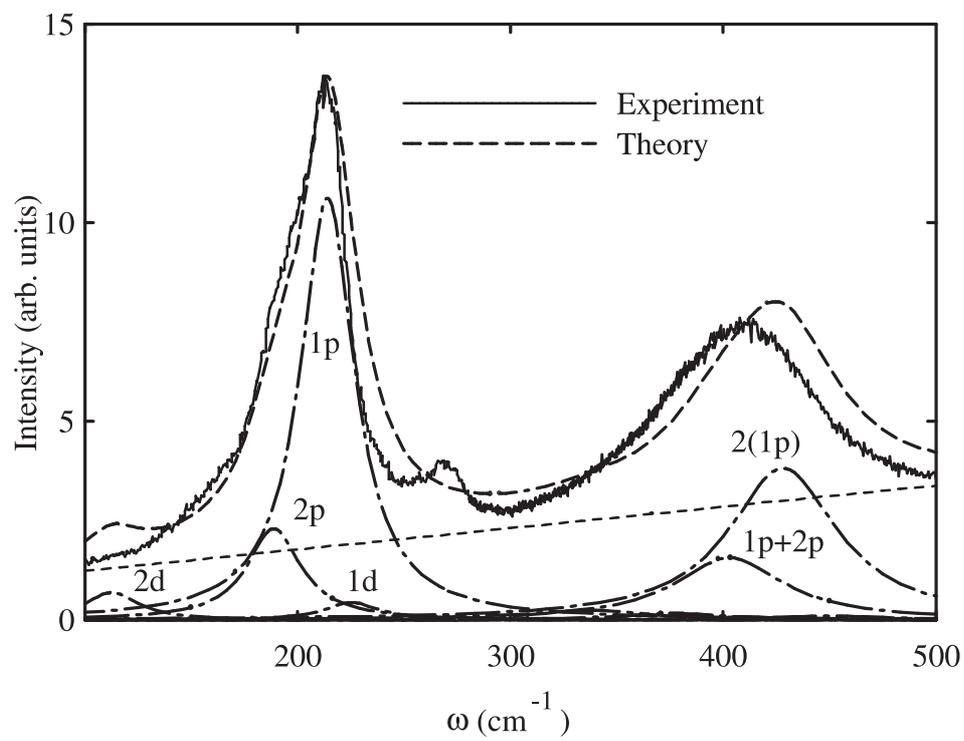
\noindent Fig. 2. Calculated Raman scattering spectrum (the heavy dashed
curve) for the ensemble of PbS nanocrystals with $\left\langle
R\right\rangle =1.5$ nm. The experimental spectrum of Ref. \cite{Krauss2} is
shown by the solid line. The following parameter values are used: $%
E_{g}=0.307$ eV \cite{MPZ64}; $\varepsilon _{1}\left( \infty \right) =18.5,$ 
$\varepsilon _{1}\left( 0\right) =190$ \cite{Kartheuser}.

\end{document}